\documentclass[aps,prc,twocolumn,showpacs,floatfix,superscriptaddress]{revtex4}

\usepackage{amsmath}
\usepackage{graphicx}
\usepackage{amssymb}
\usepackage[colorlinks]{hyperref}
\usepackage{subfigure}
\usepackage{sidecap}

\providecommand{\aap}{Astron.Astrophys.}
\providecommand{\physrep}{Phys. Rep.}

\begin{document}

\title{Neutrino-pair emission from nuclear de-excitation in core-collapse
supernova simulations}

\author{T.~Fischer}
\affiliation{Institute for Theoretical Physics, University of Wroc{\l}aw, pl. M. Borna 9, 50-204 Wroc{\l}aw, Poland} 

\author{K.~Langanke}
\affiliation{GSI Helmholtzzentrum f\"ur Schwerioneneforschung,
  Planckstra{\ss}e~1, 64291 Darmstadt, Germany}
\affiliation{Institut f{\"u}r Kernphysik, Technische Universit{\"a}t
  Darmstadt, Schlossgartenstra{\ss}e 2, 64289 Darmstadt, Germany}
\affiliation{Frankfurt Institute for Advanced Studies, Ruth-Moufang Str.,
D-   Frankfurt, Germany}

\author{G.~Mart{\'i}nez-Pinedo}
\affiliation{Institut f{\"u}r Kernphysik, Technische Universit{\"a}t
  Darmstadt, Schlossgartenstra{\ss}e 2, 64289 Darmstadt, Germany} 
\affiliation{GSI Helmholtzzentrum f\"ur Schwerioneneforschung,
  Planckstra{\ss}e~1, 64291 Darmstadt, Germany}

\begin{abstract}
  We study the impact of neutrino-pair production from the
  de-excitation of highly excited heavy nuclei on core-collapse
  supernova simulations, following the evolution up to several 100~ms
  after core bounce.  Our study is based on the AGILE-Boltztran
  supernova code, which features general relativistic radiation
  hydrodynamics and accurate three-flavor Boltzmann neutrino transport
  in spherical symmetry.  In our simulations the nuclear de-excitation
  process is described in two different ways.  At first we follow the
  approach proposed by Fuller and Meyer~[Astrophys. J. \textbf{376},
  701 (1991)], which is based on strength functions derived in the
  framework of the nuclear Fermi-gas model of non-interacting
  nucleons.  Secondly, we parametrize the allowed and forbidden
  strength distributions in accordance with measurements for selected
  nuclear ground states.  We determine the de-excitation strength by
  applying the Brink hypothesis and detailed balance.  For both
  approaches, we find that nuclear de-excitation has no effect on the
  supernova dynamics.  However, we find that nuclear de-excitation is
  the leading source for the production of electron antineutrinos as
  well as heavy-lepton flavor (anti)neutrinos during the collapse
  phase. At sufficiently high densities, the associated neutrino
  spectra are influenced by interactions with the surrounding matter,
  making proper simulations of neutrino transport important for the
  determination of the neutrino-energy loss rate.  We find that even
  including nuclear de-excitations, the energy loss during the
  collapse phase is overwhelmingly dominated by electron neutrinos
  produced by electron captures.
\end{abstract}

\date{\today}

\maketitle

\section{Introduction}

Massive stars end their lifes as supernovae, triggered by the collapse
of their central core. It has long been recognized that neutrinos play
a crucial role for the dynamics of the collapsing
core~\cite{Bethe90,Janka:2007} and the associated supernova
nucleosynthesis~\cite{Woosley:2002zz}. Once the electron chemical potential gets
sufficiently large at densities of order $10^9$~g~cm$^{-3}$, electrons
are captured on protons bound in nuclei~\cite{Langanke:2007ua}. This
has the following two important consequences for the collapse
dynamics, it reduces the pressure which the relativistic electron gas
can supply against the collapse, and the (electron) neutrinos, which
are produced from weak-interaction processes, leave the star carrying
away energy and lepton number. In fact, this cooling mechanism keeps
the core at relatively low entropies so that heavy nuclei survive the
collapse and are the dominant component of the nuclear
composition \cite{Bethe78}. 
With increasing core density neutrino interactions with
matter become growingly more relevant. Coherent elastic neutrino
scattering on nuclei~\cite{Langanke.Martinez-Pinedo:2003} leads to
neutrino trapping for densities in excess of about
$10^{12}$~g~cm$^{-3}$. Inelastic neutrino scattering on electrons, and
to a lesser extent on nuclei~\cite{Langanke:2007ua},
down-scatters neutrinos in energy and ultimately leads to the
thermalization of the trapped neutrinos. Hence in the late stage of
the collapse i.e. at densities in excess of $10^{12}$~g~cm$^{-3}$, a
Fermi sea of electron neutrinos is formed in the inner core that
effectively Pauli-blocks further electron captures.

\begin{figure}[h]
\subfigure[]{
\includegraphics[width=0.375\columnwidth]{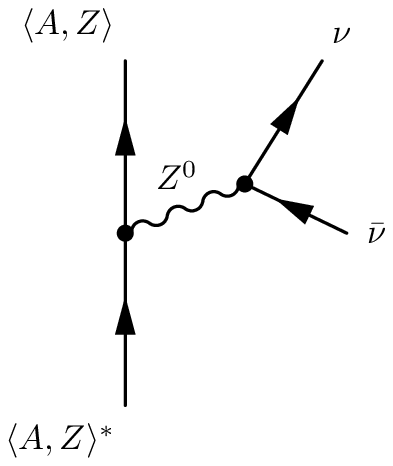}
\label{fig:pair-scheme}}
\hfill
\subfigure[]{
\includegraphics[width=0.4\columnwidth]{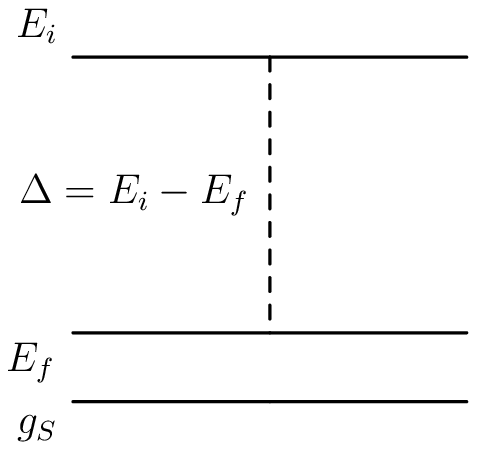}
\label{fig:levels}}
\caption{Feynman diagram for the de-excitation of a heavy nucleus via the
emission of a neutrino pair in graph~(a).
Schematic diagram in graph~(b), for the situation of an excited heavy nucleus
with excitation energy $E_i$ decaying to a state of lower energy $E_f$, above
the ground state $g_S$.
The quantity $\triangle$ is the energy difference between initial and final
states.}
\end{figure}

In supernovae weak charged-current reactions produce electron
neutrinos and antineutrinos. Other neutrino types can only be
generated by processes governed by neutral current; i.e. in the form
of neutrino-antineutrino pairs.  In current supernova simulations,
neutrino-pair production is considered via electron-positron
annihilation, nucleon-nucleon bremsstrahlung, and the annihilation of
trapped electron-neutrino and antineutrino pairs into heavy-lepton
flavor neutrino-antineutrino pairs~\cite{Buras:2002wt}.  
It has been argued that the de-excitation of highly
excited nuclei can be the dominant neutrino-pair producing process in
the hot environment of the collapsing core~\cite{Fuller:1991}
(for illustration, see Fig.~\ref{fig:pair-scheme}). As the presence of
the electron neutrino sea does not block the production of muon and
tau neutrino-antineutrino pairs as well as that of electron
antineutrinos, simultaneously produced in a pair with a high-energy
electron neutrino, nuclear de-excitation might further reduce the
entropy of the collapsing core if the neutrinos produced by the
process will be able to leave the core during the dynamical timescale
of the collapse. Indeed if the produced neutrinos have energies low
enough to leave the stellar core, it is speculated that the de-excitation
process `likely acts as a thermostat for the collapsing core'; i.e. in
a self-regulating process more escaping neutrinos are being produced
the hotter the core temperature~\cite{Fuller:1991}.

In this manuscript, we report on supernova simulations which for the
first time include the nuclear de-excitation processes. To this end we
evolve a 11.2~M$_\odot$ star~\cite{Woosley:2002zz} from the
presupernova progenitor through the core collapse, bounce and
post-bounce evolution for more than 300~ms. Our study is based on the
AGILE-Boltztran supernova code, see ref.~\cite{Liebendoerfer:2004} and
references therein for additional details.

In nuclear de-excitation a highly excited state at energy, $E_i$,
decays via $Z^0$ emission to a final state at lower energy, $E_f$ (for
illustration, see Fig.~\ref{fig:pair-scheme}).  The energy difference,
$\Delta=E_i-E_f$, between the nuclear states is shared by the $\nu
\bar{\nu}$-pair which is created by the decay of the $Z^0$ boson.
(The situation is illustrated in Fig.~\ref{fig:levels}.)  For small
values of $\Delta$, the total nuclear strength is dominated by
Gamow-Teller transitions.  At higher energy differences $\Delta$,
forbidden transitions will contribute.  Modeling such transitions for
an ensemble of thermally excited nuclear states at moderate
temperatures ($T \approx 1-2$ MeV) typical for supernova environment
is a challenging problem. It cannot be solved by current models for
the relevant heavy nuclei, due to the extremely high density of states
involved. As a first approach, rates for nuclear excitation and
de-excitation have been estimated on the basis of the nuclear Fermi
gas model; i.e. describing the nucleus as an ensemble of
non-interacting nucleons occupying a set of shell model
orbits~\cite{Fuller:1991}.  The rates for up-transitions, i.e. the
nuclear excitation via the absorption of a neutrino pair, obtained in
this model have been recently qualitatively supported based on the
interacting shell model~\cite{Wendell:2013}. 
As our first scheme to describe the de-excitation 
process in a supernova simulation, we have adopted the respective 
rates of ref~\cite{Fuller:1991}
and have incorporated them into the supernova AGILE-Boltztran code. 

To study the sensitivity of the supernova results on potential
uncertainties in the deexcitation rates
we describe the latter also in an alternative
approach.  Here we start from the observation that the relevant
Gamow-Teller (and forbidden) strength is constrained experimentally
for selected nuclear ground states showing that the various strengths
is mainly concentrated in strong transitions (the GT and spin-dipole
giant resonances).  Adopting Brink's hypothesis~\cite{Brink:1955}, which assumes that the transition strengths on excited
nuclear states is the same as for the ground state, and exploiting the
principle of detailed balance allows us to derive the `downwards'
transition strengths required for modeling the de-excitation rate from
experimentally motivated parametrization of the Gamow-Teller and
forbidden ground state strengths.

The paper is organized as follows.  In sec.~\ref{sec:core-coll-supern}
we review our core-collapse supernova model, and in
sec.~\ref{sec:impr-weak-proc} we discuss the implementation of new
rates for electron-captures on heavy nuclei as well as the
neutrino-pair-production from nuclear de-excitation.  In
sec.~\ref{sec:simul-results-stell}, we apply these new weak rates in
core-collapse simulations and discuss the results. We
close the manuscript with a summary in sec.~\ref{sec:summary-conclusions}.

\section{Core-collapse supernova model}
\label{sec:core-coll-supern}

Our core collapse supernova code, AGILE-Boltztran, is based on general
relativistic radiation hydrodynamics and three-flavor Boltzmann
neutrino transport in spherical symmetry (for details, see
ref.~\cite{Liebendoerfer:2004} and references therein). The set of
weak processes that we consider in our supernova simulations are
listed in Table~\ref{table-nu-reactions}, including the
references which we have used. 
The charged-current processes~(1) and (2) as well as the
elastic scattering reaction~(4) are important in regions where the
composition is dominated by nucleons. The charged-current reactions
with nucleons are treated in the zero-momentum transfer
approximation~\cite{Bruenn:1985en,Reddy:1998}.
Electron-positron annihilation~(7)
dominates the production of neutrino pairs at low and intermediate
densities, while nucleon-nucleon Bremsstrahlung becomes important at
higher densities, around a tenth of normal nuclear matter density.
We also include the annihilation of trapped electron neutrino pairs (process (9) in
Table~\ref{table-nu-reactions}). Inelastic neutrino scattering on
electrons and positrons, reaction~(6) in
Table~\ref{table-nu-reactions}, is the dominant thermalization process
for neutrinos~\cite{Langanke:2007ua}.

\begin{table}[ht!]
\centering
\caption{Neutrino reactions considered, including references.}
\begin{tabular}{ccc}
\hline
\hline
$\;\;\;\;\;\;\;\;$
&
Weak process\footnote{Note:
  $\nu=\{\nu_e,\bar{\nu}_e,\nu_{\mu/\tau},\bar{\nu}_{\mu/\tau}\}$ 
and $N=\{n,p\}$}
&
References
\\
\hline
1 & $e^- + p \rightleftarrows n + \nu_e$ & \cite{Reddy:1998} \\ 
2 & $e^+ + n \rightleftarrows p + \bar\nu_e$ & \cite{Reddy:1998} \\
3 & $e^- + (A,Z) \rightleftarrows (A,Z-1) + \nu_e$ & \cite{Juodagalvis:2010} \\
4 & $\nu + N \rightleftarrows \nu' + N$ & \cite{Bruenn:1985en,Mezzacappa:1993gm} \\
5 & $\nu + (A,Z) \rightleftarrows \nu' + (A,Z)$ & \cite{Bruenn:1985en,Mezzacappa:1993gm} \\
6 & $\nu + e^\pm \rightleftarrows \nu' + e^\pm$ & \cite{Bruenn:1985en,Mezzacappa:1993gx} \\
7 & $e^- + e^+ \rightleftarrows  \nu + \bar{\nu}$ & \cite{Bruenn:1985en} \\
8 & $N + N \rightleftarrows  \nu + \bar{\nu} + N + N $ & \cite{Hannestad:1997gc} \\
\hline
\end{tabular}
\label{table-nu-reactions}
\end{table}

At the high density environment of a core-collapse supernova, 
electron neutrinos are more strongly coupled to
matter by process~(1) than electron antineutrinos by process (2) (see
Table~\ref{table-nu-reactions}). Hence, electron antineutrinos decouple at
generally higher densities than electron neutrinos. Moreover,
heavy-lepton neutrinos do not interact by charged-current processes
at the relevant conditions and are 
hence even less coupled to matter than electron antineutrinos;
they decouple from matter at even higher densities. Consequently,
during the post-bounce mass accretion phase when matter at neutrino
decoupling is extremely neutron rich, the following hierarchy holds
for the average neutrino energies $\langle E_{\nu_{\mu/\tau}} \rangle
\gtrsim \langle E_{\bar\nu_e} \rangle > \langle E_{\nu_e}
\rangle$~\cite{Raffelt:2001,Keil:2003,Fischer:2012}. 

The equation of state (EoS) in core-collapse supernova studies has to
handle a variety of conditions that relate to different nuclear
regimes. It spans from isospin-symmetric matter at low densities and
temperatures dominated by heavy nuclei up to supersaturation densities
where matter is extremely neutron rich and temperatures reach several
tens of MeV. AGILE-Boltztran uses a flexible EoS module that allows
for the use of a large variety of currently available supernova
EoS~\cite{Lattimer:1991nc,Shen:1998gg,Hempel:2009mc}. For a comparison
study of different supernova EoS, see
refs.~\cite{Hempel:2012,Steiner:2013}. For the current study, we apply
the EoS from ref.~\cite{Lattimer:1991nc} with compressibility
$K=220$~MeV for matter at temperatures above $T=0.45$~MeV. The baryon
EoS is then computed based on the ideal gas approximation. On top of
the baryons, contributions from electrons, positrons and photons are
added~\cite{Timmes:1999}.

\section{Weak processes with heavy nuclei}
\label{sec:impr-weak-proc}

\subsection{Electron captures on heavy nuclei}

Compared to previous studies performed with the AGILE-Boltztran code
we have improved the description of electron capture by implementing
the reaction rates for electron captures on nuclei (process~(3) in
Table~\ref{table-nu-reactions}) as derived in
ref.~\cite{Juodagalvis:2010}.  The authors of
ref.~\cite{Juodagalvis:2010} determined their rate tabulation from
individual capture rates for about 3000 nuclei, valid for matter in
nuclear statistical equilibrium (NSE).  To derive the individual rates
the authors adopted a hierarchy of nuclear models to ensure the
appropriate description of the electron capture process at all
conditions of the collapse.  In particular, the capture rate for
nuclei in the mass range $A=45-64$ was derived on the basis of the
interacting shell model.  It guarantees an accurate and detailed
description of the allowed strength distribution~\cite{Cole:2012}
required to describe the electron capture rate at the moderate density
conditions at which these medium-mass nuclei dominate.  The capture
rate for heavier nuclei with $A=65-120$ were derived within the
framework of the Random Phase Approximation (RPA) with partial nuclear
occupation numbers obtained in large-scale Shell Model Monte Carlo
(SMMC) calculations at finite temperature \cite{Koonin97,Langanke:2003ii,Hix:2003}. 
These two data sets have
been supplemented by individual rates for more than 2000 additional
nuclei using an SMMC+RPA approach, similar to the one introduced in
ref.~\cite{Langanke:2001} but with parametrized occupation numbers.

The authors of ref.~\cite{Juodagalvis:2010} provide a table for the
electron-capture rates and neutrino spectra for a fixed 3-dimensional
grid in temperature $T$, electron fraction $Y_e$, and density $\rho$,
valid for NSE above $T>0.45$~MeV.  We apply a linear interpolation
scheme in these three variables to determine the rates and spectra for
the appropriate astrophysical conditions.

\subsection{Neutrino-antineutrino pair emission and absorption}

Emission and absorption of neutrino pairs is a process that can
potentially affect the dynamics of the core. In this section we
derive expressions for the rate of heavy-nuclei de-excitation and
excitation that can be used in core-collapse supernova simulations.
The decay rate of a nucleus from a state with excitation energy $E_1$
via emission of neutrino-antineutrino pairs of a particular flavor is
given by~\cite{Fuller:1991}:
\begin{widetext}
\begin{eqnarray}
\lambda_{\nu\bar\nu}(E_1) &=&
\frac{2 \pi}{\hbar}
\frac{G_F^2 g_A^2 (4 \pi)^2}{(2\pi \hbar c)^6}
\int_0^{E_1}  dE_2
\int_0^\Delta E_{\bar\nu}^2 E_\nu^2 dE_{\nu}
\frac{1}{(4\pi)^2} \int_{-1}^{+1} d\mu_\nu \int_{-1}^{+1} d \mu_{\bar\nu}
\label{eq:rate1}
\\
& &
[1-f_{\nu}(E_{\nu},\mu_\nu)][1-f_{\bar\nu}(E_{\bar\nu},\mu_{\bar\nu})]
\int_0^{2\pi}d\varphi_\nu
\int_0^{2\pi}d\varphi_{\bar\nu}
S^{\text{down}}(E_1,\Delta,\cos\theta),
\nonumber
\end{eqnarray}
\end{widetext}
where
\begin{equation}
\Delta = E_1 - E_2 = E_\nu + E_{\bar\nu},
\label{eq:delta}
\end{equation}
and $G_F$ is the Fermi coupling constant and $g_A$ the weak axial-vector
coupling constant. 
We account for the presence of neutrinos by including the neutrino and antineutrino
distribution functions $f_\nu(E_\nu,\mu_\nu)$ that due to our assumption of spherical
symmetry depend only on the radius, energy and the cosine of the angle with respect
to the radial direction, $\mu_\nu$, and not in the azimuthal angle, $\varphi_\nu$.
$S^{\text{down}}(E_1,\Delta,\cos\theta)$ is the strength function connecting a state
with excitation energy $E_1$ to a state with excitation energy $E_2=E_1 -\Delta$
and $\theta$ the angle between the emitted neutrino-antineutrino pair that is
related to the neutrino angles by:
\begin{eqnarray}
\cos\theta = \mu_\nu \mu_{\bar{\nu}} +
\sqrt{(1-\mu_\nu^2)(1-\mu^2_{\bar\nu})} \cos\phi,
\\
(\phi\equiv\varphi_\nu - \varphi_{\bar\nu})\,\,\,.
\nonumber
\end{eqnarray}
The first integral of equation~(\ref{eq:rate1}) represents the contributions of all the
different states to which the excited state can decay.
Furthermore, we have included a factor of $(4\pi)^2$ in front of the integrals
that represents the value of the angular integrals assuming isotropicity.
Here, we follow the standard convection in nuclear beta decay.
We can perform a change of variables, using equation~(\ref{eq:delta}), so that the integrals
are performed as a function of the neutrino and antineutrino energies:
%
\begin{eqnarray}
\label{eq:rateEnu}
&&\lambda_{\nu\bar\nu}(E_1) =
\lambda_0
\int_0^{E_1}  dE_\nu \int_{-1}^{+1} d\mu_\nu E_\nu^2 [1-f_{\nu}(E_{\nu},\mu_\nu)] 
\\
&&\times
\int_0^{E_1-E_\nu} dE_{\bar\nu} \int_{-1}^{+1} d \mu_{\bar\nu} E_{\bar\nu}^2
[1-f_{\bar\nu}(E_{\bar\nu},\mu_{\bar\nu})] 
\nonumber
\\ 
&&\times
\frac{1}{(4\pi)^2}
\int_0^{2\pi}d\varphi_\nu
\int_0^{2\pi}d\varphi_{\bar\nu}
S^{\text{down}}(E_1,E_\nu+E_{\bar\nu},\cos\theta),
\nonumber  
\end{eqnarray}
%

where 
\begin{equation}
\lambda_0 = \frac{G_F^2 g_A^2}{2 \pi^3\hbar (\hbar c)^6}\,\,.
\end{equation}
In order to obtain the total rate of neutrino-pair production from heavy-nuclei de-excitations,
all possible transitions have to be included properly weighted by the appropriate Boltzmann
factor and the level-density, $\rho(E,J)$.
One then obtains:
\begin{equation}
\lambda_{\nu\bar\nu} = \frac{1}{Z(T)}  
\int_0^\infty dE (2J+1) \rho(E,J) \lambda_{\nu\bar\nu}(E) e^{-E/T}
\label{eq:rate_total_level_density}
\end{equation}
with $Z(T)=\int dE (2J+1) \rho(E,J) e^{-E/T}$, the partition function.
Using equation~(\ref{eq:rateEnu}) and after changing the integration order we obtain:
%
\begin{eqnarray}
  \label{eq:ratetotal}
&&\lambda_{\nu\bar\nu} =\lambda_0
\int_0^{\infty}  dE_\nu \int_{-1}^{+1} d\mu_\nu  E_\nu^2 [1-f_{\nu}(E_{\nu},\mu_\nu)]  
\\
&&
\times
\int_0^{\infty} dE_{\bar\nu} \int_{-1}^{+1} d \mu_{\bar\nu}
 E_{\bar\nu}^2
[1-f_{\bar\nu}(E_{\bar\nu},\mu_{\bar\nu})] 
\nonumber
\\
& &
\times
\frac{1}{(4\pi)^2}
\int_0^{2\pi}d\varphi_\nu
\int_0^{2\pi}d\varphi_{\bar\nu}
\,\mathcal{S}^{\text{emi}}(T,E_\nu+E_{\bar\nu},\cos\theta),
\nonumber  
\end{eqnarray}
%
where we have introduced the thermal strength function for emission of
a neutrino-antineutrino pair:
%
\begin{eqnarray}
\label{eq:sthermal}
&&
\mathcal{S}^{\text{emi}}(T,E_\nu+E_{\bar\nu},\cos\theta) = \frac{1}{Z(T)} \int_0^\infty
dE
\\
&&
\times
(2J+1) \rho(E,J) \, S^{\text{down}}(E,E_\nu+E_{\bar\nu},\cos\theta) e^{-E/T} \,\,.
\nonumber
\end{eqnarray}
%
The strength function, which  connects states 1 and 2 ($S^{\text{down}}$) with 
the energy relation
$E_1 = E_2 +\Delta$, is associated to the strength function connecting states 2 and 1
($S^{\text{up}}$) by detailed balance~\cite{Thomas:1964,Thomas:1968,Grover.Gilat:1967}: 
%
\begin{eqnarray}
\label{eq:detbal}
(2J_1+1) \rho(E_1,J_1)
S^{\text{down}}(E_1,\Delta,\cos\theta) =
\\
= (2 J_2 + 1) \rho(E_2,J_2)  S^{\text{up}}(E_2,\Delta,\cos\theta).  
\nonumber
\end{eqnarray}
%
Note that the above expression is commonly used to relate the down and
up gamma-ray strength functions in the calculation of radioactive
capture reactions~\cite{Capote.Herman.ea:2009}.  We obtain the
following relationship between the thermal strength functions for
emission and absorption of a pair of neutrinos:
%
\begin{eqnarray}
\label{eq:sthermaldb}
\mathcal{S}^{\text{emi}}(T,E_\nu+E_{\bar\nu},\cos\theta) &=&
\mathcal{S}^{\text{abs}}(T,E_\nu+E_{\bar\nu},\cos\theta) \,\,\,\,\,\,\,\,\,\,
\\
&\times&
\exp\left(-\frac{E_\nu + E_{\bar\nu}}{T}\right),
\nonumber 
\end{eqnarray}
%
where $\mathcal{S}^{\text{abs}}$ is related to $S^{\text{up}}$ by an
equation similar to equation~(\ref{eq:sthermal}).

The total number of neutrino pairs produced per unit of volume and time,
$\Lambda_{\nu\bar\nu}$, is given 
by the decay rate for nuclear species $i$, called
$\lambda^i_{\nu\bar\nu}$, weighted 
with the total number
density of nuclei $n_i$ and summing over all nuclear species present in the medium,
\begin{equation}
  \label{eq:lambdaall}
  \Lambda_{\nu\bar\nu} = \sum_i n_i \lambda^i_{\nu\bar\nu}
\end{equation}
In the Boltzmann representation of neutrino transport this quantity is
normally described by the pair-emission kernel,
$R^{\text{emi}}(E_\nu+E_{\bar\nu},\cos\theta)$.  The total rate for
$\nu\bar\nu$-emission, based on the reaction kernel $R^{\text{emi}}$,
has the general form,
%
\begin{eqnarray}
\label{eq:ratekernel}
\Lambda_{\nu\bar\nu} &=& \frac{1}{(2\pi\hbar c)^6}
\int_0^{\infty} dE_\nu E_\nu^2 
\int_{-1}^{+1} d\mu_\nu 
\left(1-f_{\nu}(E_{\nu},\mu_\nu)\right)
\nonumber
\\
&&
\times
\int_0^{\infty} dE_{\bar\nu} E_{\bar\nu}^2
\int_{-1}^{+1} d \mu_{\bar\nu} 
\left(1-f_{\bar\nu}(E_{\bar\nu},\mu_{\bar\nu})\right)
\nonumber
\\
&&
\times
\int_0^{2\pi} d\varphi_\nu
\int_0^{2\pi} d\varphi_{\bar\nu}  
\,\,R^{\text{emi}}(E_\nu+E_{\bar\nu},\cos\theta).
\end{eqnarray}
%
Comparing this expression with equation~(\ref{eq:ratetotal}), we obtain:
\begin{equation}
  \label{eq:kernelstrength}
  R^{\text{emi}}(\Delta,\cos\theta) = \frac{2\pi}{\hbar}
  G_F^2 g_A^2 \sum_i n_i \mathcal{S}^{\text{emi}}_i(T,\Delta,\cos\theta).
\end{equation}
The absorption kernel is related to absorption strength by a similar
expression. As a consequence of equation~\ref{eq:sthermaldb}, we
obtain the general detailed balance expression for the
kernel~\cite{Bruenn:1985en}:

\begin{eqnarray}
  \label{eq:kerneldb}
  R^{\text{emi}}(E_\nu+E_{\bar\nu},\cos\theta) &=&
  R^{\text{abs}}(E_\nu+E_{\bar\nu},\cos\theta) \\
  & &\times \exp\left(-\frac{E_\nu + E_{\bar\nu}}{T}\right).  \nonumber 
\end{eqnarray}
Finally, the de-excitation process is considered in the Boltzmann transport
equation~\cite{Bruenn:1985en,Rampp:2002,Fischer:2012} 
by adding an appropriate contribution to the source term:
\begin{widetext}
\begin{eqnarray}
\label{eq:boltzmanncol}
B_{\text{NDE}} [f_\nu]  &=&  \frac{1}{c (2\pi\hbar c)^3} \left\{
\left[1-f_{\nu}(E_{\nu},\mu_\nu)\right]
\int_0^\infty dE_{\bar\nu} E_{\bar\nu}^2
\int_{-1}^{+1} d \mu_{\bar\nu} 
 \left[1-f_{\bar\nu}(E_{\bar\nu},\mu_{\bar\nu})\right]
\int_0^{2\pi} d\varphi_{\bar\nu}
R^{\text{emi}}(E_\nu+E_{\bar\nu},\cos\theta) \right.
\nonumber
\\
&&
\left. -
f_{\nu}(E_{\nu},\mu_\nu)
\int_0^\infty dE_{\bar\nu} E_{\bar\nu}^2
\int_{-1}^{+1} d \mu_{\bar\nu} 
f_{\bar\nu}(E_{\bar\nu},\mu_{\bar\nu})
\int_0^{2\pi} d\varphi_{\bar\nu}
R^{\text{abs}}(E_\nu+E_{\bar\nu},\cos\theta)\right\},
\end{eqnarray}
\end{widetext}
with a similar equation for $f_{\bar\nu}$, i.e. $f_\nu \leftrightarrow
f_{\bar\nu}$. For the particular case in which neutrinos escape
freely from the stellar core we can define the total (including all
neutrino flavors) nuclear de-excitation rate as:
\begin{equation}
  \label{eq:decayrate}
  \lambda = \frac{3 G_F^2 g_A^2}{60\pi^3\hbar (\hbar c)^6}
  \int_0^\infty  E^5 \bar{\mathcal{S}}^{\text{emi}}(T,E) dE.
\end{equation}
The factor 3 accounts for the three possible neutrino flavors that can
be produced in the decay and we have used the fact that the strength
function depends only on the sum of neutrino energies to perform one
of the energy integrations in equation~(\ref{eq:ratetotal}).  We have
also introduced the angle averaged thermal strength function:
%
\begin{eqnarray}
\label{eq:Sangleaver}
\bar{\mathcal{S}}(T,E) &=& \frac{1}{(4\pi)^2} 
\int_{-1}^{+1} d\mu_\nu   \int_0^{2\pi}d\varphi_\nu
\int_{-1}^{+1} d \mu_{\bar\nu} \\
&&
\times\int_0^{2\pi}d\varphi_{\bar\nu}
\mathcal{S}(T,E,\cos\theta)
\nonumber
\end{eqnarray}
%
Equivalently we can define the de-excitation energy loss rate:
\begin{equation}
\label{eq:eloss}
\dot{Q} = \frac{3 G_F^2 g_A^2}{60\pi^3\hbar (\hbar c)^6}
\int_0^\infty  E^6 \bar{\mathcal{S}}^{\text{emi}}(T,E) dE.
\end{equation}
We can also define the neutrino (antineutrino) spectra, i.e. the
number of neutrinos per energy and time, produced by nuclear
de-excitations by integrating over the antineutrino (neutrino) energy
as follows:
\begin{equation}
\label{eq:4}
\mathcal{N}_\nu (E) = \frac{G_F^2 g_A^2}{2\pi^3\hbar (\hbar c)^6}
E^2 \int_0^\infty d E_{\bar\nu} E^2_{\bar\nu}
\bar{\mathcal{S}}^{\text{emi}}(T,E_\nu+E_{\bar\nu}) 
\end{equation}

\subsection{Strength function}
\label{sec:strength-function}

To determine the neutrino-pair de-excitation rate we have to determine
the temperature, neutrino pair energy and angle dependence of the
thermal strength function $\mathcal{S}(T,E,\cos\theta)$, see
Eq.~(\ref{eq:sthermal}).  As the process is expected to be relevant at
temperatures around 1~MeV and higher, correspondingly at nuclear excitation
energies above $\sim 10$~MeV, a state by state evaluation of the total rate
is prohibited due to the overwhelmingly large density of levels
involved.  Hence one has to turn to an `averaged' way in describing
the respective strength function.  Here we will follow two alternative
approaches.

In the first approach we follow the proposal of Fuller and
Meyer~\cite{Fuller:1991} which developed analytical expressions for
the emission and absorption thermal strength functions using a Fermi
gas model.  The parameters where adjusted to reproduce the results
obtained in an independent single-particle shell model.  Fuller and
Meyer considered a strength distribution that consists of both allowed
and first-forbidden contributions and assumed different angular
dependence for each of them:
%
\begin{eqnarray}
\label{eq:sallowforb}
\mathcal{S}(T,E,\cos\theta) &=& \mathcal{S}_A(T,E) P_A(\cos\theta)
\\
&+& \mathcal{S}_F(T,E) P_F(\cos\theta).
\nonumber
\end{eqnarray}
%
We refer to the work of Fuller and Meyer~\cite{Fuller:1991}
for the particular form of the strength
functions, $S_A$ and $S_F$, and their dependence on nuclear mass and charge,
$A$ and $Z$ respectively.
However, we used the angular dependence of the functions $P_A$ and $P_F$ as:
\begin{equation}
\label{eq:3}
P_A(\cos\theta) = 1-\frac{1}{3}\cos\theta, \quad P_F(\cos\theta) = 1
\end{equation}
which is differently normalized as defined by Fuller and Meyer and has been
chosen such that:
\begin{equation}
\label{eq:2}
\bar{\mathcal{S}}^{\text{emi}}(T,E) = \mathcal{S}_A(T,E) + \mathcal{S}_F(T,E),
\end{equation}
which is the angle-averaged thermal strength function, see equation~(\ref{eq:Sangleaver}).
With the angular dependence of equation~(\ref{eq:sallowforb}), the azimuthal integral
in the source term for NDE can be performed to obtain: 
\begin{widetext}
\begin{eqnarray}
\label{eq:boltzmanncol2}
B_{\text{NDE}}[f_\nu] = \frac{2\pi}{c (2\pi\hbar c)^3}
&\biggl\{&
\left[1-f_{\nu}(E_{\nu},\mu_\nu)\right]
\int_0^\infty dE_{\bar\nu} E_{\bar\nu}^2
\int_{-1}^{+1} d \mu_{\bar\nu} 
\left[1-f_{\bar\nu}(E_{\bar\nu},\mu_{\bar\nu})\right]
\mathcal{R}^{\text{emi}}(E_\nu+E_{\bar\nu},\mu_\nu,\mu_{\bar\nu})
\nonumber
\\
& &-
f_{\nu}(E_{\nu},\mu_\nu)
\int_0^\infty dE_{\bar\nu} E_{\bar\nu}^2
\int_{-1}^{+1} d \mu_{\bar\nu}
f_{\bar\nu}(E_{\bar\nu},\mu_{\bar\nu})
\mathcal{R}^{\text{abs}}(E_\nu+E_{\bar\nu},\mu_\nu,\mu_{\bar\nu})
\biggr\},
\end{eqnarray}
with
\begin{equation}
\label{eq:5}
\mathcal{R}(E,\mu_\nu,\mu_{\bar\nu}) = \frac{2\pi}{\hbar} G_F^2
g_A^2 \sum_i n_i \left[\mathcal{S}^i_A(T,E)
\left(1-\frac{\mu_\nu\mu_{\bar\nu}}{3}\right) + \mathcal{S}^i_F(T,E)\right]
\end{equation}
\end{widetext}

In a recent work, Wendell~Misch~\emph{et~al}~\cite{Wendell:2013} have
calculated allowed down strength functions for several nuclei
including $^{28}$Si, $^{47}$Ti and $^{56}$Fe at different excitation
energies based on the diagonalization shell-model.  In their study
they approximate the thermal strength to the one obtained for an
excitation energy equivalent to the average thermal excitation energy
$\langle E \rangle$ determined assuming a Fermi gas model, $\langle E
\rangle = T^2 A/8$.  This approach has two main disadvantages.  First,
it violates detailed balance according to
equation~(\ref{eq:sthermaldb}) and more importantly for the
calculation of de-excitation rates it results in a sharp cutoff in the
production of neutrino pairs of energies larger than the thermal
average energy.  The production of these high-energy neutrinos is
suppressed by the Boltzmann factor but it is favored by the large
phase space dependence, see eq.~(\ref{eq:decayrate}).

In the following, we propose an alternative approach which fulfills
detailed balance by construction and accounts for the production of
neutrinos with energies greater than the average thermal excitation.
We derive the thermal strength guided by experimental knowledge of the
allowed and forbidden strengths for nuclear ground states.  It is well
known that the allowed (Gamow-Teller) and forbidden (dipole) strength
$S^{\text{up}}$ on nuclear ground states resides mainly in giant
resonances \cite{Osterfeld:1991}.  
This is observed in ($p,p^\prime$) \cite{Hausser:1987,Uchida:2003}
(as well as in
charge-exchange experiments of $N=Z$ nuclei \cite{Anderson:1991,Zegers:2011}, 
which determine
$S^{\text{up}}$ due to isospin symmetry).  Much information about giant
resonances has been obtained from ($e,e^\prime$) experiments performed
over the entire nuclear chart \cite{Heyde:2010}.  
The experiments have been supplemented
by theoretical studies where in particular large-scale shell model
calculations give a very fair account of the detailed structure of the
allowed strength distribution \cite{Zuker:2005}, 
while studies of the forbidden strength
is the domain of models like the (Quasiparticle) Random Phase
Approximation \cite{Kolbe:2003}.  
The experimental and theoretical studies indicate that
the allowed strength is concentrated in a giant resonance with a
centroid energy at about $E_x=8$--10~MeV, reflecting mainly an
excitation of nucleons between spin-orbit partner orbitals.  As the
dipole strength corresponds to a transition between two adjacent major
shells, its giant resonance resides at somewhat higher excitation
energies with a centroid around $E_x=18$--24~MeV.  Both the giant
Gamow-Teller and dipole resonances are strongly fragmented over many
states and the strength distribution can be approximated by Gaussian
distribution around the respective centroids \cite{Haxton:1996,Qian:1996}.

We will assume that the giant resonances built on excited states are
located at the same relative excitation energy as for the ground
state.  This is commonly known as Brink hypothesis and implies that
the up strength at any excitation energy is equal to the one of the
ground state,
$S^{\text{up}}(E,\Delta)=S^{\text{up}}_{\text{gs}}(\Delta)$.  This
approximation is commonly done in calculations of astrophysical
reaction rates based on the statistical
model~\cite{Litvinova.Loens.ea:2009} and implies that the thermal
absorption strength becomes independent of the temperature and is
given by:
\begin{equation}
  \label{eq:sbrink}
  \mathcal{S}^{\text{abs}}(T,\Delta) = S^{\text{up}}_{\text{gs}}(\Delta) 
\end{equation}
However, we note that due to nuclear structure effects (pairing,
angular momentum mismatch etc.)  and the low density of levels, the
upward strengths on the ground state vanish or are strongly suppressed
at low excitation energies.  This behavior is not expected for the
upward strength on excited states as they are populated at the
temperatures of interest for the neutrino pair de-excitation process.
We account for this expectation by assuming that the strength is
fragmented over a larger energy range than for the ground state, while
the total strength is the same as for the ground state.  Guided by
these assumptions we make the following ansatz for the absorption
thermal strength:
\begin{equation}
\label{eq:Sup_gauss}
S^{\text{abs}}(\Delta) = S_A g(\Delta,\mu_A,\sigma_A) + S_F
g(\Delta,\mu_F,\sigma_F) 
\end{equation}
where $S_A$ and $S_F$ are the total allowed and forbidden strength and
$g$ is the normalized strength distribution with centroid $\mu$ and
standard deviation $\sigma$ that we assume to follow the Gaussian
distribution.  The total allowed strength, $S_A = 5$, is chosen in
accordance with the value found for nuclei in the iron
region~\cite{Langanke.Martinez-Pinedo.ea:2004,Fearick.Hartung.ea:2003,Juodagalvis.Langanke.ea:2005},
while the forbidden strength, $S_F = 7$, is chosen guided by RPA
calculations~\cite{Juodagalvis.Langanke.ea:2005,Kolbe.Langanke:2001}.
For the centroid and standard deviation we use $\mu_A = 9$~MeV,
$\sigma_A = 5$~MeV, $\mu_F = 22$~MeV and $\sigma_F = 7$~MeV.  The
emission thermal strength is then obtained by applying detailed
balance, see eq.~(\ref{eq:sthermaldb}). Our ansatz should be
considered as a simple approximation as it neglects a temperature
dependence of the width parameters and assumes the strength to be the
same for all nuclei.  At low temperatures we also expect deviations
from a Gaussian distribution caused by the discrete level structure
for the allowed strength and the non-equilibration of parity in the
level density at low energies for the forbidden strength.
\begin{figure*}[htb]
\centering\includegraphics[width=.68\columnwidth]{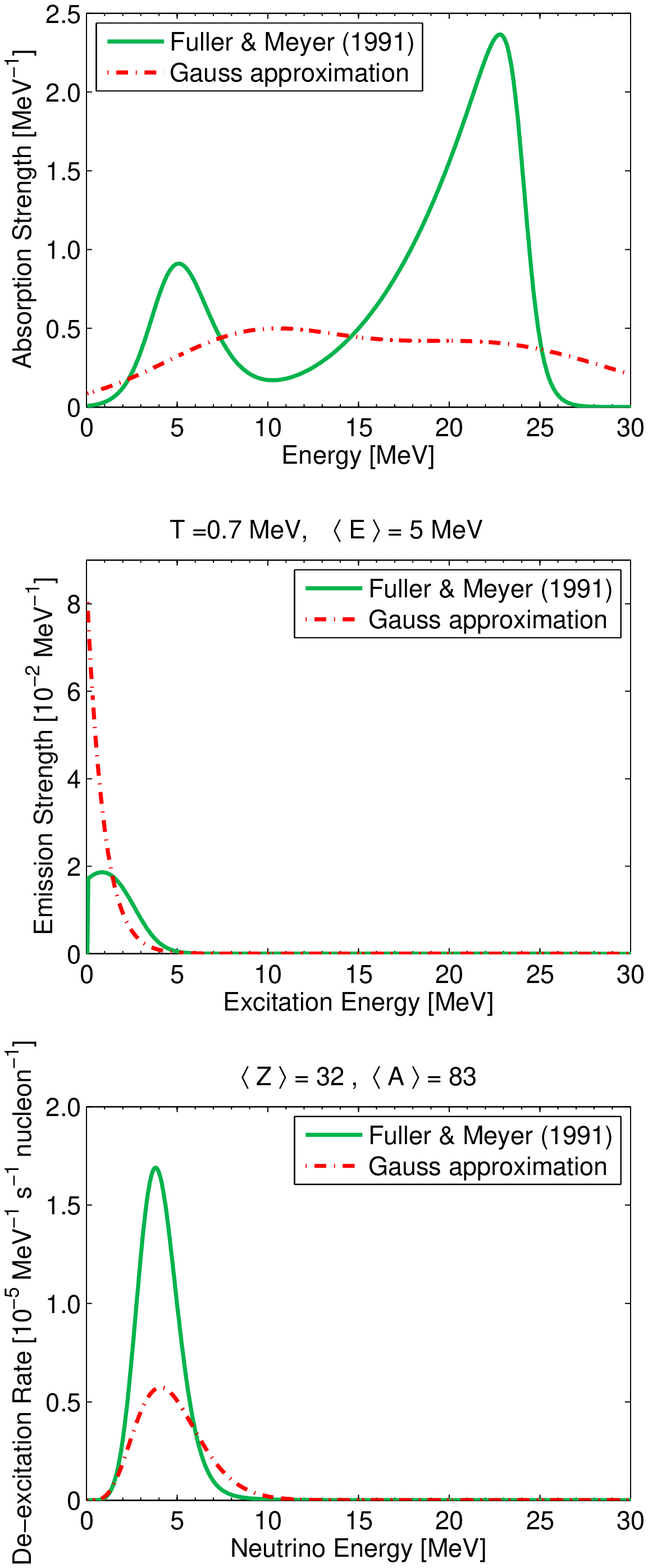}
\includegraphics[width=.68\columnwidth]{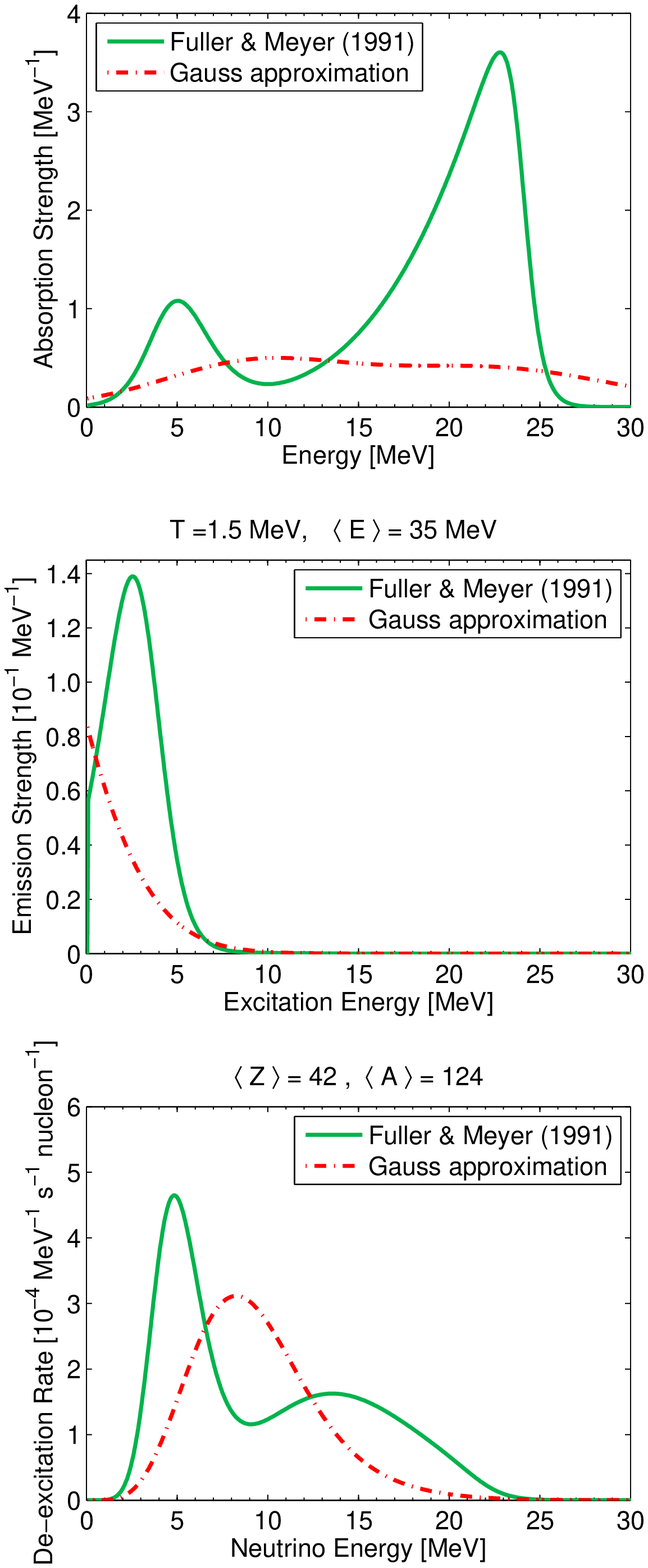}
\includegraphics[width=.68\columnwidth]{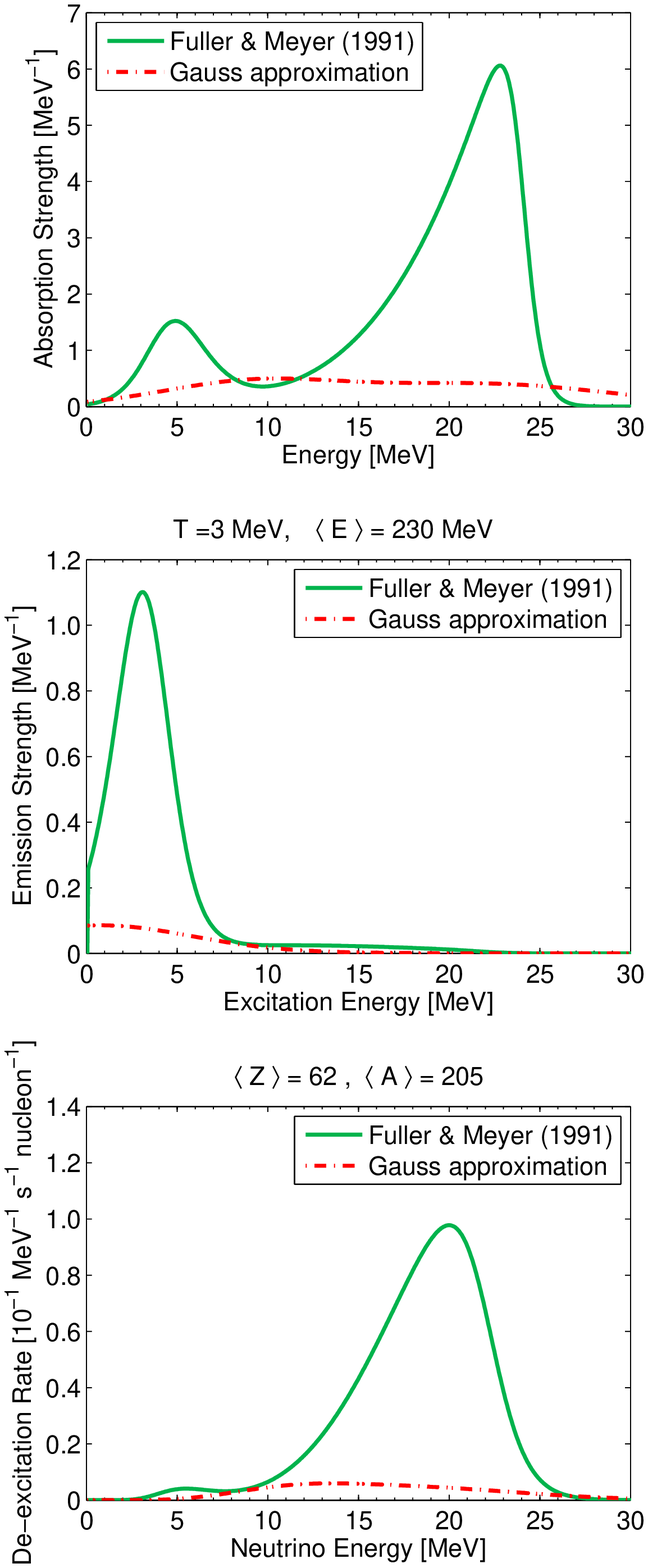}
\caption{Strength for absorption (top panels) and emission (middle
  panels) of neutrino pairs based on the Fermi-gas model of
  ref.~\cite{Fuller:1991} (green lines) and our approach (red lines,
  see text for a description) at temperatures of $T=0.7$~MeV (left
  panel), $T=1.5$~MeV (middle panel), and $T=3.0$~MeV (right panel).
  The corresponding conditions are listed in table~\ref{table-lumin}.
  The bottom panels show the de-excitation rate versus energy, i.e.
  the integrand in Eq.~(\ref{eq:decayrate}).}
\label{fig-spectra}
\end{figure*}

The top panels of Fig.~\ref{fig-spectra} compare the absorption
strength function $\mathcal{S}^{\text{abs}}$ (see
equation.~\ref{eq:sthermaldb}) for the two approaches considered in
our study. We use the composition given by the EoS at selected
temperatures, densities and $Y_e$ values obtained from our
simulations.  The strength function derived by Fuller and Meyer shows
rather distinct peaks for allowed and forbidden transitions, while in
our ansatz the strengths for these transitions are fragmented over a
wider energy range.  We note that the differences become quite
pronounced at low ($E < 1$~MeV) and high ($E>25$~MeV) energies where
the strength suggested by Fuller and Meyer basically vanishes, while
the Gaussian ansatz, expression~\ref{eq:Sup_gauss}, shows noticeable
strength. Furthermore, the Fuller and Meyer approach \cite{Fuller:1991}
predicts a total strength that is proportional to the number of nucleons.
This explains the increase of strength with temperature, as observed
in Fig. \ref{fig-spectra}, as the nuclear
composition moves to heavier nuclei with increasing temperature
in the supernova environment. Hence the two cases considered
here can be understood as extreme cases for the absorption strength.
  
The middle panels of Fig.~\ref{fig-spectra} shows the thermal emission
strength functions $\mathcal{S}^{\text{emi}}$, which are obtained from
the thermal absorption functions by multiplication with the Boltzmann
factor $\exp(-E/T)$.  Due to this factor the thermal emission
strength is strongly depending on temperature.  Note that while our
absorption strength is independent on temperature and is the same for all
nuclei, the one suggested by Fuller and Meyer is weakly depending on temperature
and increases with increasing nucleon number.  We note that the
differences in the two absorption strength functions lead to quite
strong deviations in the emission functions. The emission functions
derived from the Gaussian absorption function shows a continuous
decrease with energy reflecting mainly the exponential decrease of the
Boltzmann factor as the corresponding absorption function varies only
slowly in the energy range of importance.  As the absorption strength
suggested by Fuller and Meyer has very little strength at vanishing
energy, this energy range is also suppressed in the emission strength
functions, which show a pronounced peak at moderately low energies.
We note that there is also emission strength at energies above the
thermal average energy (denoted by $\langle E \rangle$ in the middle
panels for each temperature), due to thermal population of states at
higher energies.  However, the Boltzmann factor forces the emission
strength to vanish at high energies.  Obviously the developing tail of
the emission functions depend on temperature and on the assumptions on
the absorption strength function.

The tails in the emission functions are more pronounced in the
de-excitation rate, i.e.  the integrand of
equation~(\ref{eq:decayrate}).  This quantity is shown in the bottom
panels of Fig.~\ref{fig-spectra}.  One clearly observes that the
production of neutrinos with energies larger than the average nuclear
excitation energy is important.  In particular at low temperatures,
$T=0.7$ MeV, where $\langle E \rangle \sim 5$~MeV, the maximum of the
pair de-excitation rate is located at higher energies than the average
excitation energy. Comparing the results for the two strength function
approaches at $T=1.5$ MeV, one finds that the Fermi-gas
approximation~\cite{Fuller:1991} exhibits two peaks which can be
associated with the distinct allowed and forbidden transitions.  At
higher temperatures (see the right panel of Fig.~\ref{fig-spectra}),
forbidden transitions even dominate the de-excitation rate. The
emission strength resulting from the Gaussian strength function shows
a single peak structure at all temperatures due to the rather broad
strength functions used in our parametrization. However, it also
predict an increasing role of forbidden strength with increasing
temperature. 

\subsection{Spectra comparison under supernova conditions}

\begin{figure}[htb]
\centering
\includegraphics[width=1\columnwidth]{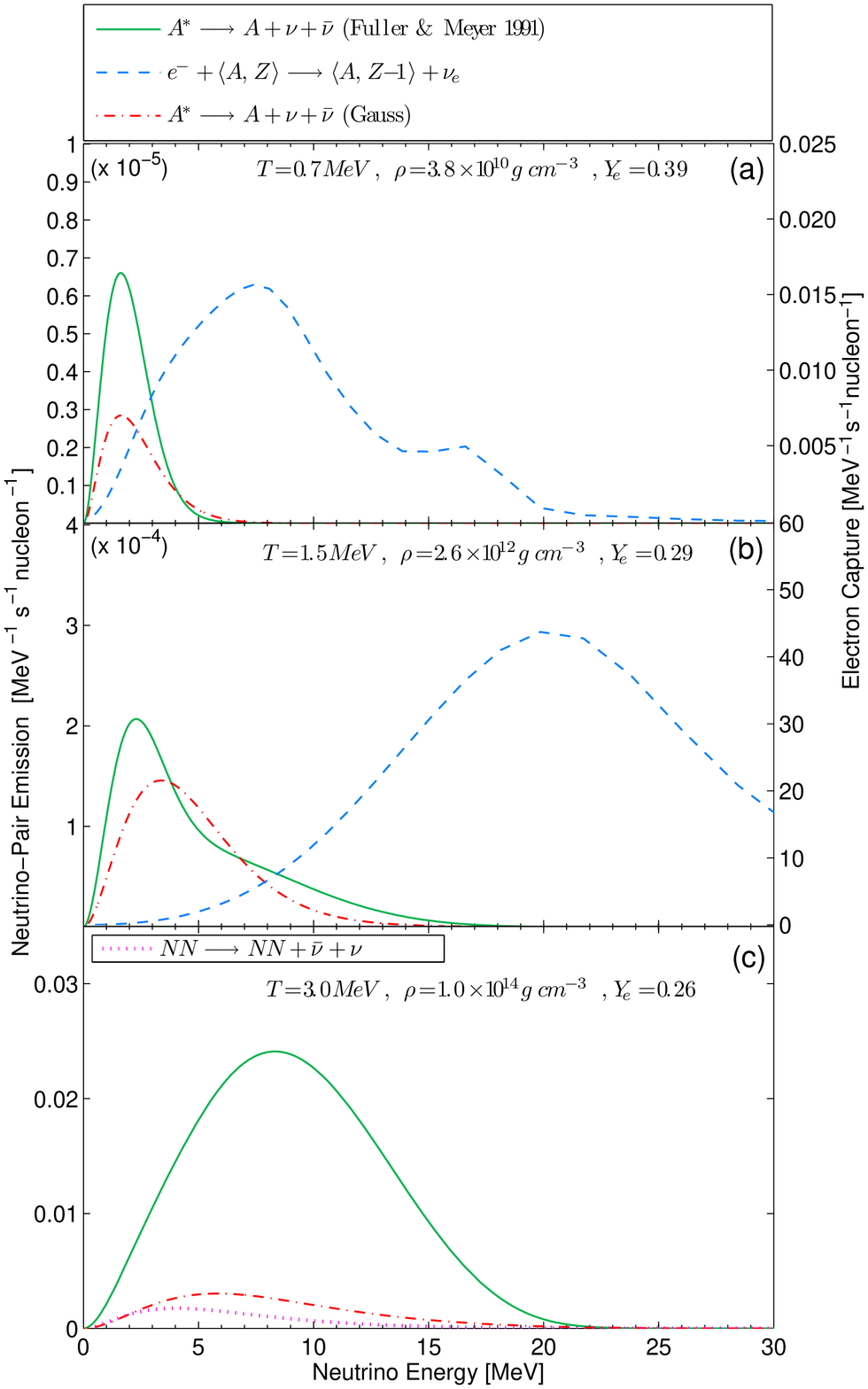}
\caption{Total neutrino emission spectra comparing electron captures
  on heavy nuclei~\cite{Juodagalvis:2010} (red dash-dotted lines) and
  heavy-nuclei de-excitation based on Fuller and Meyer~\cite{Fuller:1991}
  (blue solid lines) as well as
  expression~(\ref{eq:Sup_gauss}) (red dash-dotted lines), at low (a),
  intermediate (b) and high temperatures/densities (c), i.e. 
  corresponding to entries in table~\ref{table-lumin}. In the bottom
  panel, instead of electron captures we show the
  $\nu\bar\nu$-emission rate from $N-N-$Bremsstrahlung (magenta dotted
  line).  (color version online). For the upper and middle panel the
  neutrino emission spectra have to be multiplied by the number shown
  in parenthesis.}
\label{fig-pair}
\end{figure}

Figure~\ref{fig-pair} shows the (anti)neutrino spectra produced by
nuclear de-excitation (as defined in \ref{eq:4}) and electron capture.
The spectra are compared at three different
stages of the collapse phase, where we have chosen the temperatures to
match those adopted in Fig.~\ref{fig-spectra}. We note that the
spectra do not include final-state Pauli blocking of the
neutrinos. Hence the de-excitation spectra basically depend on
temperature, with a density dependence arising for the nuclear Fermi
gas results with its A-dependent strength function due to the change
of the nuclear composition. As the electron Fermi energy strongly
depends on density, the neutrino spectra produced by electron capture
do as well. The three conditions chosen correspond to the progenitor
phase (upper panel), the stage of electron neutrino trapping (middle
panel) and the late phase before bounce where 
a transition to uniform nuclear matter is taking
place.  At all conditions, electron captures produce neutrino spectra
with significantly higher energies.  This is confirmed in
Table~\ref{table-lumin} in which we have summarized the average
neutrino energies produced by electron capture and for our two
approaches to nuclear de-excitation, assuming that the produced
neutrinos leave the star unhindered.  To understand these results we
note that for electron capture the relevant energy scale is the
electron Fermi energy, which grows from about 10 MeV to 100 MeV in the
density range covered by Figure~\ref{fig-pair} 
\cite{Langanke.Martinez-Pinedo:2003}.  
Hence this process
produces on average higher-energy neutrinos than nuclear de-excitation
for which the average neutrino energies $\langle E \rangle \propto T$ due to the
  dominance of the Boltzmann factor $\exp(-E/T)$ as explained above
in connection with Fig.~\ref{fig-spectra}.  From Figure~\ref{fig-pair}
and Table \ref{table-lumin} we observe that at the lower temperatures
the Fermi gas based model for nuclear de-excitation produces neutrinos
with slightly lower average neutrino energies than our Gaussian
strength model, while it is the opposite for temperatures $T\gtrsim
1.5$ MeV.  As explained above this is related to the stronger
contributions of forbidden transitions in the strength function of
ref.~\cite{Fuller:1991}.

The neutrino spectra shown in Fig.~\ref{fig-pair} are unnormalized and
the most important result is obtained by comparing the scales of the
two processes, electron capture and nuclear de-excitation, showing
that the former exceeds the later by about 5 orders of magnitude.
This implies that electron captures will be the dominating
weak-interaction process for the global properties of supernovae,
while the importance of nuclear de-excitation is constrained to the
results concerning neutrino types other than electron neutrinos.
These expectations are confirmed in our supernova simulations which we
turn to in the following section. Note that in Fig.~\ref{fig-pair}~(c)
we show $N-N-$Bremsstrahlung, in addition to nuclear de-excitation,
which starts to become important as heavy nuclei disappear at
densities in excess of $\sim10^{13}$~g~cm$^{-3}$.

\begin{table}[htp]
\centering
\caption{Average energy of the neutrino produced by electron capture
  and nuclear de-excitations ($\nu\bar\nu$), for selected conditions
  during the collapse.}
\begin{tabular}{cccccc}
  \hline \hline
  $T$ (MeV) & $\rho$ (g~cm$^{-3}$) & $Y_e$ & $\langle E \rangle^{\text{ecap}}$ (MeV)&
  \multicolumn{2}{c}{$\langle E \rangle^{\nu\bar\nu}$ (MeV)} \\ \cline{5-6}
  & & & Ref.~\cite{Juodagalvis:2010} & Ref.~\cite{Fuller:1991} & This work \\
  \hline
  0.55 & $2.0\times 10^{10}$ & 0.42 &  $\,\,\,\,$8.54 &  1.82 &  1.86 \\
  0.70 & $3.8\times 10^{10}$ & 0.39 &  $\,\,\,\,$8.91 &  2.05 &  2.32 \\
  1.00 & $4.2\times 10^{11}$ & 0.33 & 12.08 &  2.57 &  3.27 \\
  1.50 & $2.6\times 10^{12}$ & 0.29 & 21.36 &  4.93 &  4.68 \\
  2.00 & $8.5\times 10^{12}$ & 0.27 & 27.51 &  7.45 &  5.96 \\
  3.00 & $2.0\times 10^{14}$ & 0.26 & 75.46\footnote{includes only the
    contribution from $e^-$ captures on protons} & 9.21 & 9.04  \\
  \hline
\end{tabular}
\label{table-lumin}
\end{table}

Table~\ref{table-lumin} lists the expected average neutrino energies
from heavy-nuclei de-excitations ($\nu\bar\nu$), assuming that all
neutrinos produced can leave the star.  At temperatures above
  1.5~MeV, corresponding to densities above $10^{12}$~g~cm$^{-3}$, the
  average neutrino energies become larger than 5~MeV, so that
down-scattering of electrons becomes relevant, implying 
substantial changes for the neutrino spectra and making proper
neutrino transport important. This expectation is confirmed by the
results of our simulations presented in the next section.
%

\section{Simulation results of stellar core collapse}
\label{sec:simul-results-stell}

In this section, we discuss results from core-collapse supernova
simulations of the 11.2~M$_\odot$ progenitor \cite{Woosley:2002zz},
focussing on the effects of heavy-nuclei de-excitation by neutrino
pair emission.  To this end, we have performed three supernova
simulations which differ only in their treatments of the neutrino pair
de-excitation process. In two simulations we include nuclear
  de-excitation following either the prescription of Fuller
  and Meyer \cite{Fuller:1991} or our Gaussian approximation model.
The third simulation serves as a control study, in which we have
switched off neutrino pair de-excitation.  We use the extended
set of electron capture rates of
Juodagalvis~et~al.~\cite{Juodagalvis:2010} in all our simulations.  We
note that our control run yields results which agree quite well with
those obtained in refs.~\cite{Langanke:2003ii,Hix:2003}, which used a
subset of the Juodagalvis capture rates (consisting of the shell model
and SMMC+RPA rates of refs.~\cite{Langanke.Martinez-Pinedo:2000,
  Langanke.Martinez-Pinedo:2001,Langanke:2003ii}), indicating that the
extension of the electron capture rate set has only small impact on
supernova simulations. However, the control simulation leads to a
noticeable lower central electron fraction due to a longer
deleptonization phase before heavy nuclei dissolve and also to a
different electron-fraction structure towards lower densities at core
bounce than obtained in studies based on the schematic description of
electron capture provided by ref.~\cite{Bruenn:1985en}.

\begin{figure}[ht!]
\centering
\includegraphics[width=1.\columnwidth]{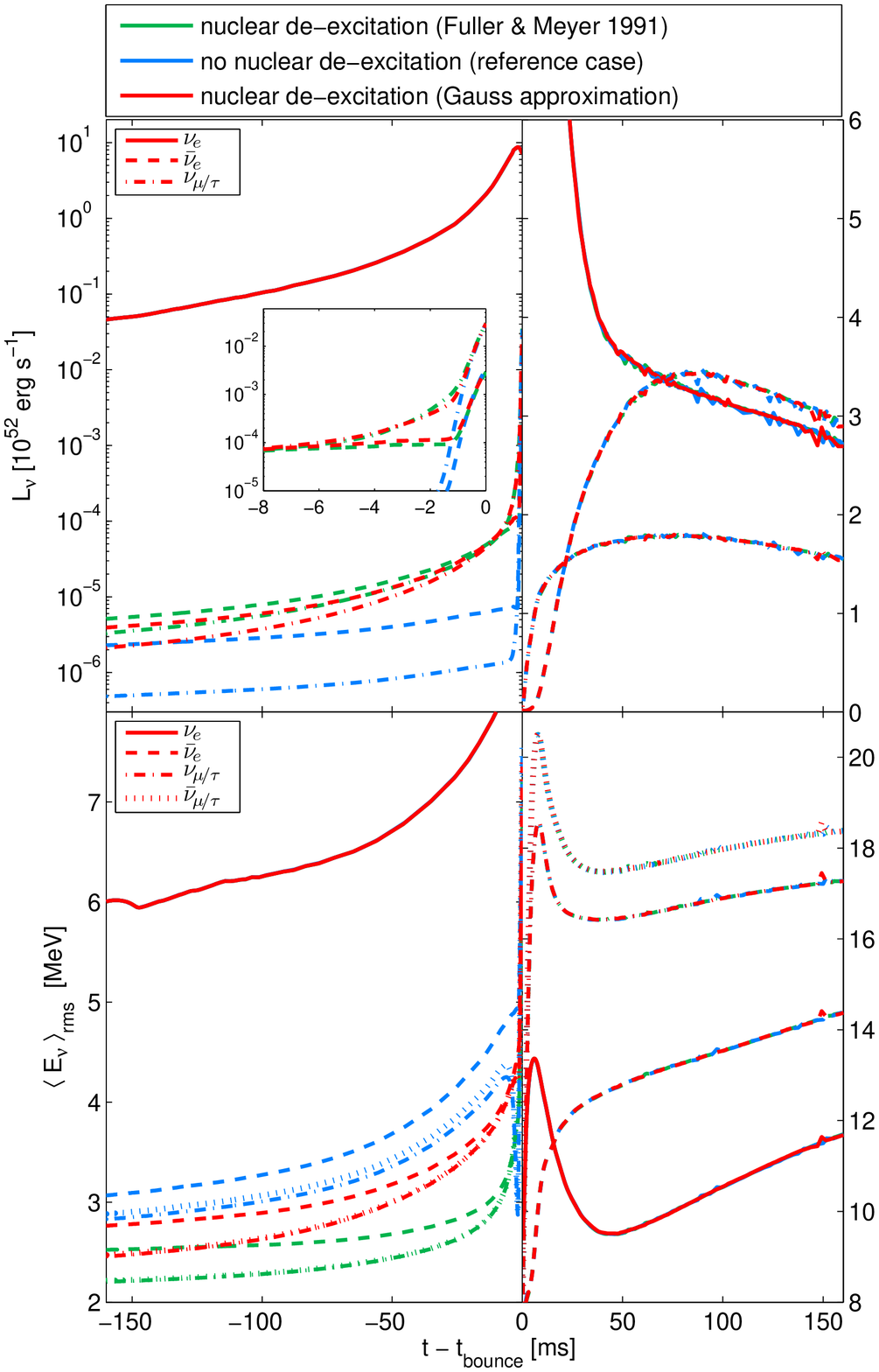}
\caption{Evolution of neutrino luminosities and root mean square
  average energies from a core-collapse supernova simulations for
  which we include the production of neutrino pairs from heavy-nuclei
  de-excitations, based on ref.~\cite{Fuller:1991} (green lines) and
  the Gauss approximation expression~(\ref{eq:Sup_gauss}) (red lines),
  in comparison to a simulation that uses the standard set of weak
  rates as listed in Table~\ref{table-nu-reactions} (blue lines) and
  otherwise identical input physics (color version online).}
\label{fig-lumin}
\end{figure}

\begin{figure}[ht!]
\centering
\includegraphics[width=1.\columnwidth]{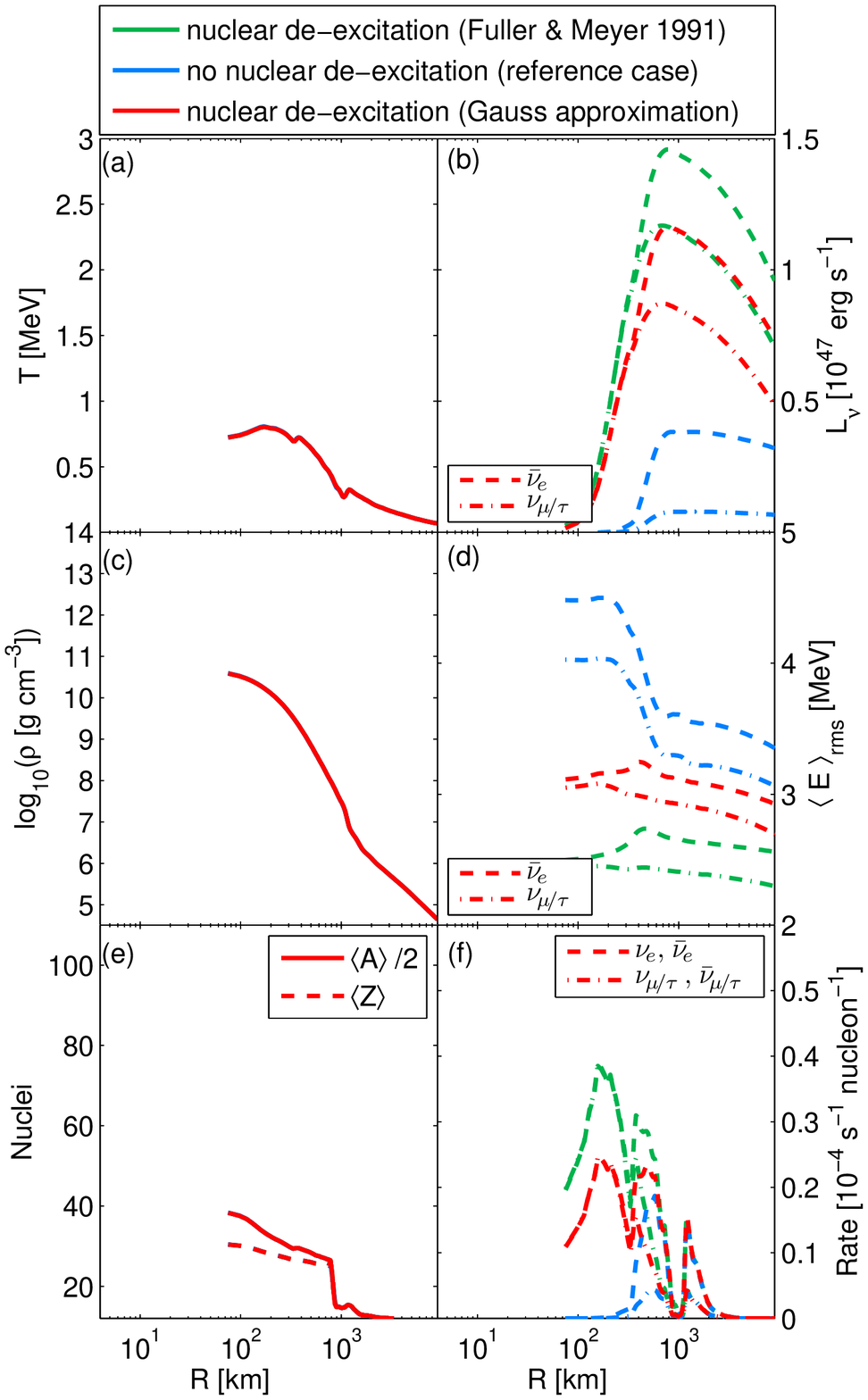}
\caption{Radial profile of selected quantities at about 50~ms before core bounce, comparing simulations for which we include the production of neutrino pairs from
heavy-nuclei de-excitations, based on ref.~\cite{Fuller:1991} (green lines), and
the Gauss approximation expression~(\ref{eq:Sup_gauss}) (red lines), in comparison
to a simulation that uses the standard set of weak rates as listed in
Table~\ref{table-nu-reactions} (blue lines) and otherwise identical
input physics (color version online).}
\label{fig-fullstate-b}
\end{figure}

\begin{figure}[ht!]
\centering
\includegraphics[width=1.\columnwidth]{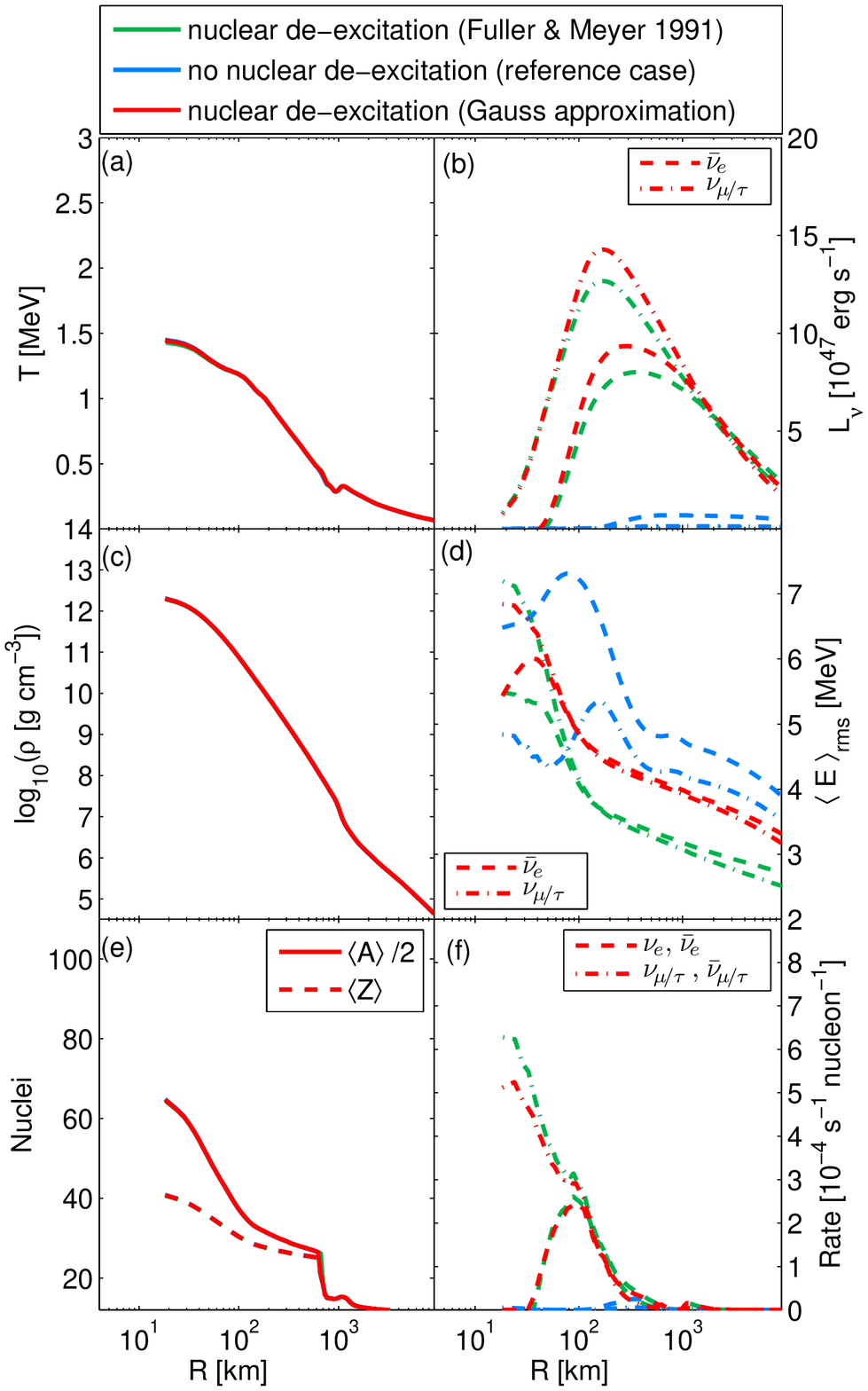}
\caption{The same configuration as Fig.~\ref{fig-fullstate-b} but at about
3.5~ms before core bounce (color version online).}
\label{fig-fullstate-d}
\end{figure}

\begin{figure}[ht!]
\centering
\includegraphics[width=1.\columnwidth]{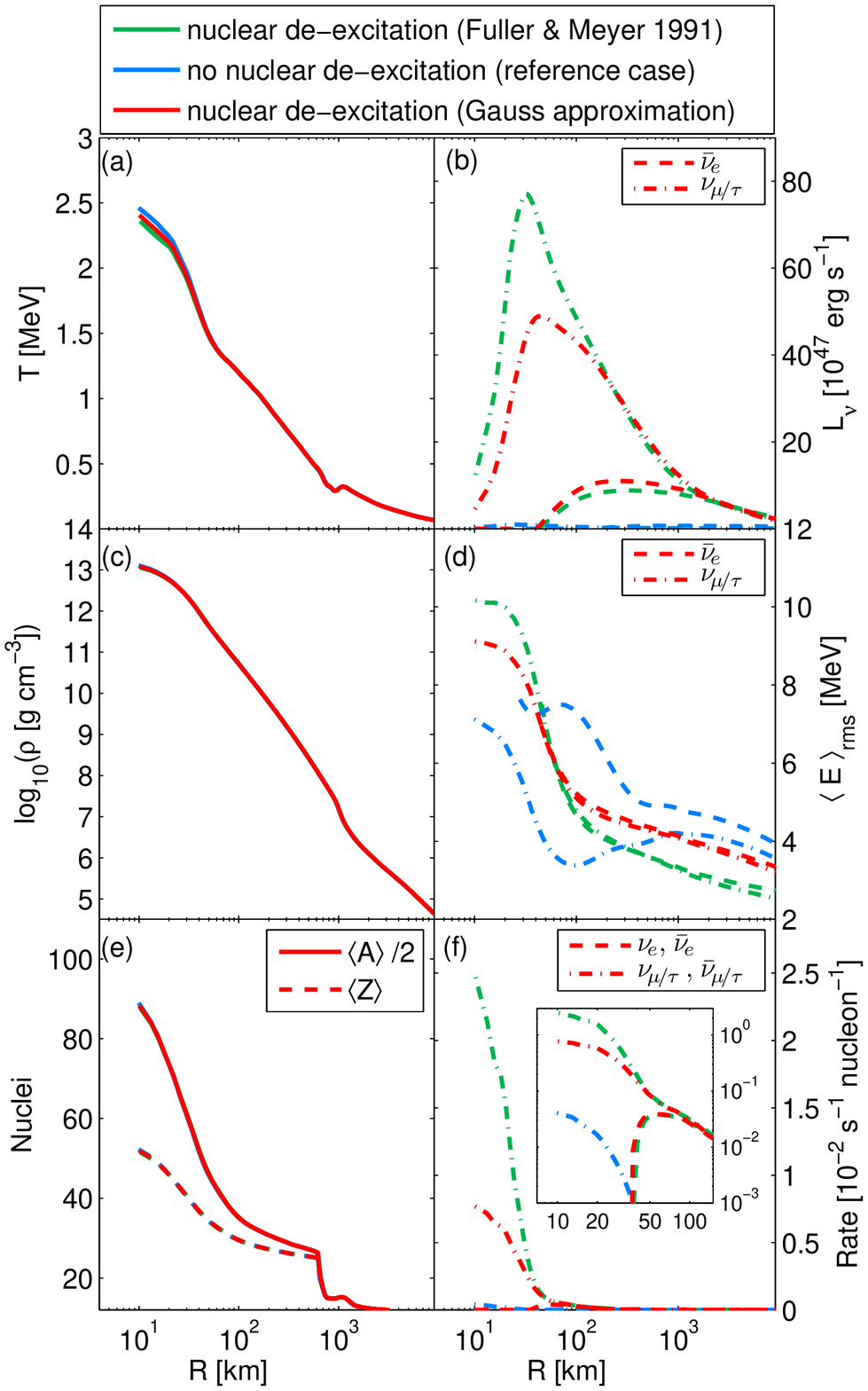}
\caption{The same configuration as Figs.~\ref{fig-fullstate-b} and
\ref{fig-fullstate-d} but at about 1.5~ms before core bounce (color version online).}
\label{fig-fullstate-e}
\end{figure}

The results of our three simulations are presented in
Fig.~\ref{fig-lumin} and in
Figs. \ref{fig-fullstate-b}-\ref{fig-fullstate-e}.
Figure~\ref{fig-lumin} shows the time evolution of the neutrino
luminosities (upper panel) and root mean square (RMS) average energies
(bottom panel) for all neutrino flavors, determined at a distance of
1000~km. Note that we determine the time of core bounce arbitrarily as
the moment when the maximum central density is reached; i.e. it is the
moment just before shock break out.  Figs.~\ref{fig-fullstate-b}-
\ref{fig-fullstate-e} show core profiles of various quantities before
bounce.

Importantly we find in these figures that the global quantities like temperature
and electron fraction profiles are the same in all three simulations (differences
seen in Fig.~\ref{fig-lumin} are due to slight mismatches in the determination of
core bounce in the different runs and due different grid resolutions).
This implies that the neutrino pair heavy nuclei de-excitation process has no impact
on the global supernova evolution.
As expected from our discussion above, the rates for electron captures on nuclei
(and protons) dominate over those for the de-excitation process.
This is confirmed in Fig.~\ref{fig-lumin} which shows that the luminosities of
electron neutrinos, arising mainly from electron capture, are about 4 orders of
magnitude larger than those of heavy flavor neutrinos during the collapse phase.

We also find in Fig.~\ref{fig-lumin} that the evolution of the $\nu_e$
luminosities and RMS average energies are the same in all three
simulations: the luminosity increases from about
$10^{51}$~erg~s$^{-1}$ to $10^{53}$~erg~s$^{-1}$ at $\nu_e$ trapping
shortly before core bounce caused by the increasing density which
increases the electron Fermi energy and the capture rates. Relatedly
the RMS average energies of $\nu_e$ neutrinos increase between $\langle
E_{\nu_e} \rangle_\text{rms}=6-10$~MeV. Note that these energies are much
lower than those of the neutrinos directly produced by electron
capture (see fig.~\ref{fig-pair}) reflecting the importance of 
down-scattering by interaction with matter.

There is, however, an important difference between electron capture
and neutrino pair heavy nuclei de-excitation which becomes noticeable
during collapse. The latter produces all neutrino types, while
electron capture is a pure source of electron neutrinos. Indeed we
find that the de-excitation process is the dominating source of
heavy-lepton flavor neutrinos and, to a lesser extent, of electron
anti-neutrinos during the collapse.  At high densities of order
$10^{13}$~g~cm$^{-3}$ also nucleon-nucleon bremsstrahlung  becomes a source of neutrinos
other than $\nu_e$.  This process is also included in our control
simulation and it is clearly visible in Fig.~\ref{fig-lumin} by the
steep rise of the $\nu_{\mu,\tau}$ and $\bar\nu_e$ luminosities at
times of 1~ms just before bounce.  In the early phase of the collapse,
i.e. at lower temperatures and densities, neutrino pair production by
electron-positron annihilation is the dominating source of
$\bar{\nu}_e$ and $\nu_{\mu,\tau}$, where, due to the charge-current
contribution the $\bar{\nu}_e$ luminosity is larger than for the heavy
neutrino flavors.  As can be seen in Fig. ~\ref{fig-pair} neutrino
pair heavy nuclei de-excitation becomes the dominating process for the
production of heavy flavor neutrinos and electron antineutrinos.  We
find rather similar $\nu_{\mu,\tau}$ luminosities during this
period for the two different approaches considered: in either case the
luminosity increases from about $10^{47}$~erg~s$^{-1}$ up to
$10^{49}$~erg~s$^{-1}$.  During the same collapse phase the average
energies of the heavy flavor neutrinos increase slightly, however,
their values $\langle E \rangle_\text{rms} \approx 2.2-3.5$~MeV are noticeably
smaller than the average energies of $\nu_e$ neutrinos.  (They are
also smaller than the average neutrino energies produced in $e^+ e^-$
annihilation that is the only heavy lepton flavor production
mechanism in our reference simulation, see discussion below.)  As
already noted above, we find slightly larger average energies when
using the Gaussian model for the strength function than for the Fuller
and Meyer approach. In either case, these values are equivalent to the
free-streaming values listed in Table~\ref{table-lumin} for core
temperatures below $\sim 1$~MeV which is the case until about 10~ms
before core bounce.  Only slightly before bounce when temperatures are
reached in excess of about 1~MeV (and
$\rho>5\times10^{11}$~g~cm$^{-3}$), the average energies start to rise
significantly and their values obtained differ from the free-streaming
values.  It points to the relevance of neutrino-matter interactions
also for heavy flavor neutrinos.


At core bounce where normal nuclear matter density is reached, heavy
nuclei dissociate into a state of homogeneous matter of nucleons and
hence the production of neutrino pairs from nuclear de-excitation
disappears.  At shock formation and during the initial shock
propagation out of the stellar core, the infalling heavy nuclei that
hit the expanding shock wave also dissociate.  Consequently at the
conditions behind the expanding shock front, weak processes are
determined by interactions with free neutrons and protons.  Hence, the
inclusion of heavy-nuclei de-excitations has no impact on the
supernova dynamics or the neutrino signal after core bounce, e.g., in
terms of the energy loss such as suggested in ref.~\cite{Fuller:1991}.
Although a small fraction of heavy nuclei exist ahead of the expanding
bounce shock before being dissociated, the conditions are such that
other pair processes dominate over the pair production from nuclear
de-excitation.  Moreover, the supernova dynamics is dominated by
charged-current processes on free nucleons behind the bounce shock in
the dissociated regime.  Consequently, neutrino pair heavy-nuclei
de-excitations has no impact in the entire post-bounce period and the
evolution of the neutrino luminosities and average energies in our
three simulation become identical.
%
%

Figs. \ref{fig-fullstate-b}-\ref{fig-fullstate-e} show core profiles
of important global quantities like temperature, density and the mean values
for the charge $\langle Z \rangle$ and mass number $\langle A \rangle$
of the nuclear composition at selected snapshots during the collapse.
Additionally we have plotted the luminosities and average energies of
the various neutrino types and the total rates of all neutrino pair
production processes (processes (7)--(9) in Table~\ref{table-nu-reactions}
and nuclear de-excitation) which are relevant
for the production of heavy flavor neutrinos and electron antineutrinos
during the collapse. Figs. \ref{fig-fullstate-b}-\ref{fig-fullstate-d}
reflect situations before and after onset of (electron) neutrino trapping
in the core with central densities $\leq 10^{11}$~g~cm$^{-3}$ and
a few $10^{12}$~g~cm$^{-3}$, respectively.
Fig. \ref{fig-fullstate-e} shows profiles close to core bounce when the
central density has reached values of a few $10^{13}$~g~cm$^{-3}$.

As our calculations involve dynamically adapting grids, comparisons between
the three calculations with and without consideration of nuclear de-excitation are
not straightforward.
We have chosen the snapshots from our three simulations such to match the core
density profile.
We then find that the profiles of the global quantities
($T, \rho, \langle Z \rangle$, $\langle A \rangle$)
are the same, independent whether nuclear de-excitation is considered in the
simulation or not, again confirming that this neutrino pair process has no influence
on the supernova dynamics.
Note the slight mismatches in the central quantities, which are due to the not
perfect matching of the evolutionary stages of the different simulations.
We further observe from the snapshot figures that, due to electron captures on
nuclei, the nuclear composition is shifted to more massive nuclei with larger neutron
excess with progressing collapse. We have used this fact already to explain the
differences in the rates and spectra calculated for the two  nuclear de-excitation
models considered in our simulation.

In our control study, without consideration of nuclear de-excitation,
$\bar\nu_e$ and $\nu_{\mu,\tau}$ are produced by electron-positron
annihilation and, at the high densities reached just before bounce in
the center, by nucleon-nucleon bremsstrahlung (see
Fig. \ref{fig-lumin}).  In Fig.~\ref{fig-fullstate-b} we observe that
the electron-positron annihilation rate is restricted to the density
regime roughly between $10^6-10^{10}$~g~cm$^{-3}$.  In this regime the
rate is proportional to the product of number densities of
electrons and positrons, $n_{e^-} n_{e^+} \sim \mu_e^3 T^3
\exp\{-\mu_e/T\}$.  As the electron chemical potential $\mu_e$, which
is roughly proportional to the third root of the density, grows faster
during the collapse than the temperature, the exponential factor
throttles the pair production by $e^+ e^-$ annihilation.  Due to the
charged current contribution the production rate of electron neutrino
pairs is larger than the one of the other two flavors.  Below
$10^{10}$~g~cm$^{-3}$, nuclear de-excitation increases the production
rate of neutrino pairs, where the relative importance is larger for
the heavy flavor neutrinos than for electron neutrinos, due to the
larger $e^+ e^-$ production rate of the latter.  For densities in
excess of $10^{10}$~g~cm$^{-3}$ the production of heavy flavor and
electron antineutrinos is basically only due to nuclear de-excitation.
The pattern of the rate follows closely the one of temperature (the
relevant parameter for the nuclear de-excitation rate which is
proportional to $T^6$). The decrease of temperature towards the center
is a consequence of the cooling by weak processes and it is
is the origin of the associated decrease of the deexcitation rate.
  The average
  neutrino energies produced by electron-positron annihilation shows a
  strong increase by about 1~MeV at a core radius of a few 100 km.
  This rise correlates with the strong change in temperature.
  However, in this range the $e^+ e^-$ rate is strongly
  suppressed by the exponential factor, and neutrinos produced by
  nuclear de-excitation dominate.  These neutrinos have smaller
  average energies than those produced by $e^+ e^-$ annihilation.  As
  a consequence, the average neutrino energies calculated in our
  simulations with the inclusion of nuclear de-excitation are
  noticeable smaller than those found in the control calculation.
Consistent with the discussion presented in Section III, we find that
the nuclear de-excitation rate is larger using the model of Fuller and
Meyer than using the Gauss approximation model, while the average
neutrino energies are smaller.

Fig. \ref{fig-fullstate-d} shows a snapshot of the core profiles after
onset of electron neutrino trapping in the center.  Also under these
conditions electron-positron annihilation and nuclear de-excitation
are the  two important neutrino pair production processes.  As,
however, the temperature has raised significantly in the inner part
(less than $\sim$ 500 km), the relative weight of the two processes
has changed significantly due to their different dependence on
temperature.  While pair production from $e^+ e^-$ annihilation still
occurs at distances of order 1000 km (where the temperature has not
noticeably changed), the temperature rise further inside causes an
increase of the nuclear de-excitation rate by more than an order of
magnitude (note the change of scales between
Figs.~\ref{fig-fullstate-b} and \ref{fig-fullstate-d}).
This makes $e^+ e^-$ negligible for the determination of the spectra of neutrinos
emitted from the core and as a consequence the average energies of the
emitted $\bar{\nu}_e$ and $\nu_{\mu,\tau}$ became almost identical
around 10~ms before bounce (see figure~\ref{fig-lumin}).
In particular, the rate for production of heavy flavor neutrinos follows the
temperature profile and increases continuously towards the center.
This is not the case for electron antineutrinos where the trapping of electron
neutrinos at densities in excess of a about $10^{11}$~g~cm$^{-3}$
hinders the production of $\nu_e \bar\nu_e$ pairs.  The trapping also
affects the spectra of $\bar\nu_e$.  The presence of trapped electron
neutrinos favors the production of $\nu_e \bar\nu_e$ pairs with low energy
antineutrinos.  Hence the respective average energies of $\bar\nu_e$
are lower than for heavy flavor neutrinos at the high energies where
electron neutrinos are trapped.  At lower densities the different
neutrino types have quite similar average neutrino energies,
reflecting the fact that nuclear de-excitation dominates as neutrino
pair production source.  Comparing the results obtained from the two
different approaches describing nuclear de-excitation, we find that
the Gaussian approximation implies slightly larger average neutrino
energies than the Fuller-Meyer ansatz, except at the highest
temperatures and densities where the tail in the neutrino spectrum, as
visible in the middle panel of Fig.~\ref{fig-pair}, leads to a
stronger increase of the average neutrino energies using the
Fuller-Meyer approach. 
Note that although the nuclear de-excitation rate increases towards
the higher densities in the center, neutrino-matter interactions affect
the transport of heavy-lepton flavor neutrinos at densities in excess of
a few $10^{11}$~g~cm$^{-3}$ (see the decreasing average energies towards 
lower densities in Fig.~\ref{fig-fullstate-d}).
As a consequence the average
energies of the neutrinos emitted from the core (see
figure~\ref{fig-lumin}) is substantially lower than the energy of
neutrinos produced in the center.
 
A snapshot of the core profiles before bounce are shown in
Fig.~\ref{fig-fullstate-e}.  The striking feature here, compared to
the other two snapshots presenting earlier collapse phases, is the
strong increase of temperature in the inner center at radii less than
about 50~km, where the densities exceed $10^{12}$~g~cm$^{-3}$.
  At these densities nucleon-nucleon
  bremsstrahlung contributes to the production of neutrino-pairs
leading to a sizable increase in the production rate of heavy flavor
neutrinos in the control simulation (see the inset in the lower right
panel of Fig.~\ref{fig-fullstate-e}).  Nevertheless the high
temperatures accelerate the production rate of heavy flavor neutrinos
by nuclear de-excitation which dominate over the $N-N$ bremsstrahlung
rate by more than an order of magnitude in this inner core.  The inset
also clearly demonstrates the suppression of electron neutrino pair
production in the density range where electron neutrinos are trapped.
As discussed above, neutrino matter interactions effects the neutrino
transport in the high-density regime, causing the peak in the
luminosity of heavy flavor neutrinos and producing a drop in
the average neutrino energies with increasing radius as neutrinos
are down scattered by scattering with electrons.  Finally we
observe that the nuclear de-excitation rate obtained for the
Fuller-Meyer approach is noticeably larger than for the Gauss
approximation which is related to the assumed $\langle A
\rangle$-dependence of the former rate and its stronger contribution
of forbidden strength.

We mention that inelastic neutrino scattering off nuclei, which is not
included in our simulations, might contribute to the
thermalization of neutrinos~\cite{Langanke:2007ua}.  For
$\nu_e$ this process increases the energy exchange with matter by
about $30\%$.  As heavy flavor neutrino scattering on electrons can
only occur by neutral current, i.e. missing the exchange term present
for electron neutrinos, the relative contribution of inelastic
neutrino scattering on nuclei is expected to be larger than for
electron neutrinos.

Can the neutrinos produced by nuclear de-excitation be observed by
neutrino detectors? As discussed above, this process is the main
source of ($\mu,\tau$) (anti)neutrinos during collapse.  However, the
associated neutrino energies are low so that neutral current reactions
are the only mean to detect these neutrino flavors by earthbound
detectors.  Moreover, as the luminosity of $\nu_e$ produced by
electron captures is several orders of magnitude larger, experimental
identification of ($\mu,\tau$) (anti)neutrinos seems impossible.  This
argument is strengthened by the fact that the $\nu_e$ average energies
are also significantly higher than those of ($\mu,\tau$)
(anti)neutrinos, prohibiting an experimental identification above a
certain energy cut.  In contrast to the core-collapse phase, this
appears to be possible during the cooling of the proto-neutron star,
where heavy-lepton flavor neutrinos have generally higher average
energies than $\nu_e$.  The low luminosity and average energy of
$\bar\nu_e$ and ($\mu,\tau$) (anti)neutrinos produced by nuclear
de-excitation, will most likely prevent direct observation.
  However, the situation might be
  different if, in case of (complete) neutrino oscillations, $\nu_e$
  and $\nu_{\mu,\tau}$ neutrinos swap their spectra~\cite{Duan:2010}.
  In such case an observation of ``$\nu_{\mu,\tau}$'' neutrinos might
  become possible based on charged-current reactions as detection
  tools.
%

\section{Summary and conclusions}
\label{sec:summary-conclusions}

The emission of neutrino-pairs from the de-excitation of highly
excited states in heavy nuclei had been proposed as a potential
additional cooling source for core-collapse supernovae.  We have
tested this suggestion by performing supernova simulations which, for
the first time, consider neutrino-pair nuclear de-excitation.  In this
article, we discussed this novel process based on two different
approaches.  First, we adopted the ansatz put forward by Fuller and
Meyer~\cite{Fuller:1991} which describes the relevant de-excitation
strength function on the basis of the Fermi-gas model of independent
nucleons.  In a second approach, we have derived the de-excitation
strength function from the inverse absorption process exploiting the
principle of detailed balance, and describing the absorption strength
function in a parametrized form guided by experimental data.  These
two choices of strength functions are quite distinct in their
predictions about the importance of forbidden contributions to the
strength, the energy centroids of the allowed and forbidden
contributions, and the dependence of the total strength on details of
the nuclear composition.  However, when incorporated in the derivation
of the neutrino-pair nuclear de-excitation rate, both approaches lead
to qualitative similar results.

Contrarily to previous expectations, we find that nuclear de-excitation have
basically no impact on the global supernova properties.
In particular, this novel weak process leaves no imprint on the dynamics
of the entire core-collapse supernova evolution up to several 100~ms post bounce.
To this end, we have performed supernova simulations in spherical symmetry based
on general relativistic radiation hydrodynamics and three-flavor Boltzmann neutrino
transport, including nuclear de-excitation for both different approaches of the
strength function.
Compared to a reference simulation, where only the standard set of weak rates is
considered, no impact on the dynamical evolution, e.g. on temperature (and entropy)
was found.
We find that electron capture on nuclei remains to be the dominating source
of energy loss through most of the infall epoch, however, producing only
electron neutrinos.

On the other hand, nuclear de-excitation produces neutrino pairs of
all flavors.  As correctly pointed out by Fuller and
Meyer~\cite{Fuller:1991}, this process produces heavy-lepton flavor
neutrinos and electron antineutrinos during the collapse phase.  In
standard supernova simulations, the production of these neutrino pairs
is governed mainly by electron-positron annihilation resulting in low
luminosities during the contraction of the stellar core.  Including
nuclear de-excitation rises their luminosities significantly but they
are still lower by several orders of magnitude than the one for
electron neutrinos produced by electron captures.  This is related to
the significantly smaller pair-production rates from nuclear
de-excitation, compared to electron captures, for most of the
core-collapse phase. Only a few milliseconds before core bounce, when
the matter temperatures reach 3--5~MeV, the strong $T$-dependence of
nuclear de-excitation leads to a strong rise of the local
pair-production rate. 
Moreover, the locally produced neutrinos from nuclear
de-excitation, in particular $(\nu_{\mu,\tau}$), interact noticeable
with matter.  These are mainly inelastic scattering on electrons (and
potentially by inelastic scattering with nuclei which has not been
considered in our simulations).  It makes the proper treatment of
neutrino transport also essential for heavy-flavor neutrinos during
the core-collapse phase.  As a result of the interactions with matter,
heavy-flavor neutrinos produced at sufficiently high densities
thermalize their spectra, similar as $\nu_e$ neutrinos, until they can
escape the collapsing stellar core.  They have lower average energies
than electron neutrinos which, together with their significantly lower
luminosities, makes their detection in absence of neutrino-flavor
oscillation scenarios very difficult.

We mention that in the present simulations inelastic neutrino scattering off
nuclei is not included.
In ref.~\cite{Langanke:2007ua} it has been argued that for $\nu_e$ this process
increases the energy exchange with matter by  about $30\%$.
For $(\mu,\nu)$ (anti)neutrinos the relevance of inelastic neutrino-nucleus
scattering might be larger, considering that the rate for inelastic neutrino
scattering on electrons is noticeably smaller for heavy flavor neutrinos than for
electron neutrinos, while inelastic neutrino-nucleus scattering rate is the same
for all flavor types.

Electron neutrinos start to become trapped at densities around
$10^{11}$~g~cm$^{-3}$, above which also the production of $\bar\nu_e$
from heavy-nuclei de-excitation sees.  Such densities are reached only
very shortly, a few milliseconds, before core bounce.  Hence the
timescale for the continued production of heavy-lepton flavor
neutrinos, which are not trapped, is not efficient enough to impact
the very final evolution until core bounce.
  At densities of a few
  times $10^{13}$~g~cm$^{-3}$, which corresponds to temperatures of
  roughly 4~MeV, other neutrino-pair processes such as nucleon-nucleon
  bremsstrahlung starts to contribute besides nuclear de-excitation.
  They produce neutrinos at significantly higher energies and at higher
  luminosity.
At even higher temperatures, when the core reaches
correspondingly nuclear matter densities and where heavy nuclei
dissolve, weak processes with free nucleons dominate.  In particular,
neutrino pairs are no longer being produced by nuclear de-excitation.
This holds for the entire post-bounce phase which is dominated by mass
accretion prior to the possible onset of an explosion, where
temperatures and entropies behind the bounce shock are so high that
heavy nuclei cannot exist.

Note that the nuclear description applied here to deduce reaction rates
for the de-excitation of heavy nuclei by the emission of neutrino-pairs is
rather crude.
Nevertheless, we do not expect that an improved treatment of nuclear
structure relevant for this weak-interaction process will significantly alter
our conclusions. 

\section*{Acknowledgment}

The supernova simulations were performed at the computer center of the GSI, 
Helmholtzzentrum f\"ur Schwerionenforschung GmbH in Darmstadt, Germany.
T.F. is supported by the Narodowe Centrum  Nauki (NCN) within the "Maestro"
program under contract No. DEC-2011/02/A/ST2/00306.
G.M.P. is partly supported by the Deutsche Forschungsgemeinschaft through
contract SFB 634, the Helmholtz International Center for FAIR within the framework
of the LOEWE program launched by the state of Hesse and the Helmholtz
Association through the Nuclear Astrophysics Virtual Institute (VH-VI-417).

\bibliography{references}
\end{document}